\documentclass[aps,prl,superscriptaddress,floatfix,reprint,showkeys,showpacs]{revtex4-1}

\usepackage{graphicx}
\usepackage{ifpdf}
\ifpdf
	\usepackage{epstopdf}
\fi

\usepackage{amsmath,amssymb,amstext}

\begin{document}
\title{Coherent magnon optics in a ferromagnetic spinor Bose-Einstein condensate}
\author{G. E. Marti}
\email{emarti@berkeley.edu}
\affiliation{Department of Physics, University of California, Berkeley, California 94720, USA}
\author{A. MacRae}
\affiliation{Department of Physics, University of California, Berkeley, California 94720, USA}
\author{R. Olf}
\affiliation{Department of Physics, University of California, Berkeley, California 94720, USA}
\author{S. Lourette}
\affiliation{Department of Physics, University of California, Berkeley, California 94720, USA}
\author{F. Fang}
\affiliation{Department of Physics, University of California, Berkeley, California 94720, USA}
\author{D. M. Stamper-Kurn}
\affiliation{Department of Physics, University of California, Berkeley, California 94720, USA}
\affiliation{Materials Sciences Division, Lawrence Berkeley National Laboratory, Berkeley, California 94720, USA}
\date{April 21, 2014}
\begin{abstract}
We measure the mass, gap, and magnetic moment of a magnon in the ferromagnetic $F=1$ spinor Bose-Einstein condensate of $^{87}$Rb. We find an unusually heavy magnon mass of $1.038(2)_\mathrm{stat}(8)_\mathrm{sys}$ times the atomic mass, as determined by interfering standing and running coherent magnon waves within the dense and trapped condensed gas. This measurement is shifted significantly from theoretical estimates. The magnon energy gap of $h\times 2.5(1)_\mathrm{stat}(2)_\mathrm{sys}\;\mathrm{Hz}$ and the effective magnetic moment of $-1.04(2)_\mathrm{stat}(8)\,\mu_\textrm{bare}$ times the atomic magnetic moment are consistent with mean-field predictions. The nonzero energy gap arises from magnetic dipole-dipole interactions.
\end{abstract}
\maketitle

Ultracold atomic systems provide access to simple, many-body models of magnetism with a unique set of experimental probes. Recent work has shown ordering towards ferromagnetic \cite{Guzman2011} and antiferromagnetic \cite{Greif2013, Islam2013} states at temperatures and coupling strengths many orders of magnitude smaller than their solid-state equivalents. Ultracold gases can realize magnetic and topological states that are otherwise difficult to access, such as those involving superfluid magnetism with multiply broken symmetries \cite{Stamper-Kurn2013}, highly symmetric $SU(N)$ magnetism \cite{Gorshkov2010, Zhang2014, Scazza2014}, and novel spin lattice models \cite{Micheli2006}. However, a unique advantage of ultracold gases may lie in their access to spin dynamics through real-time measurements of the spin density and momentum distributions. In this vein, experiments have already directly imaged a quench through a quantum critical point \cite{Sadler2006}, spin-mixing dynamics \cite{Chang2005,Liu2009}, and magnon bound states \cite{Fukuhara2013}.

In this Letter, we characterize the magnetic excitations of a spinor Bose-Einstein condensate, which is a quantum degenerate gas of bosons with a spin degree of freedom \cite{Ho1998, Ohmi1998, Stamper-Kurn2013}. The dominant spin-dependent interactions are spherically symmetric ($s$-wave), short-range collisions that are invariant under a global rotation of the spin. For a gas of ${}^{87}$Rb atoms in the $F{=}1$ hyperfine ground state, as studied here, such interactions cause the gas to be ferromagnetic at low magnetic fields, with the magnetic stiffness deriving from the kinetic-energy cost of spatial variations in the ferromagnetic superfluid order parameter.  Considering only contact interactions, this ferromagnet should support gapless magnons with an energy $E(k) = \hbar^2 k^2 / 2 m^*$ that varies quadratically with wavevector $k$ \cite{Watanabe2012}.  According to mean-field theory, the magnon mass $m^*$ equals the atomic mass $m$ \cite{Ho1998, Ohmi1998}.

We present two main results.  First, we demonstrate a form of atom interferometry based on the coherent propagation of magnons in an optically trapped spinor Bose-Einstein condensate: a magnon contrast interferometer.  Magnon waves are imprinted optically onto the condensate, allowed to propagate, and then detected with spin-dependent \emph{in situ} imaging.  If magnons are truly gapless and quadratically dispersing with $m^*=m$, then it follows that they propagate within the volume of an equilibrated trapped ferromagnetic condensate just as free atoms without the harmonic confinement of the trap \cite{Ho1998, Ohmi1998}. Our magnon contrast interferometer measures the magnon recoil frequency $E(k)/\hbar$ with a scheme similar to the atom contrast interferometer introduced in Ref.~\onlinecite{Gupta2002}, but it operates at 30-100 times lower momentum transfer and with an interaction-energy shift that is at least 100 times smaller than the mean-field shift found in its Bragg-interferometer counterpart.

Second, using contrast interferometry and other \emph{in situ} measurements of magnon dynamics, we characterize the magnon's basic properties: the excitation gap $\Delta$, the dispersion relation $E(k)$, and the magnetic moment $\mu^*$.  We find two significant differences from the predictions of a mean-field theory of $s$-wave interactions.  First, similar to solid-state examples, we find a density dependent gap in the magnon spectrum caused by magnetic dipolar interactions that, together with the anisotropy of our optical trap, invalidate the assumption of continuous rotational symmetry in our system.  Second, the magnon mass is measured to be 3.8\% larger than the bare atomic mass, a deviation that is much larger than that predicted by beyond-mean-field (Beliaev) theory \cite{Phuc2013}.

\begin{figure*}[tb!]
\begin{center}
\includegraphics[scale=1]{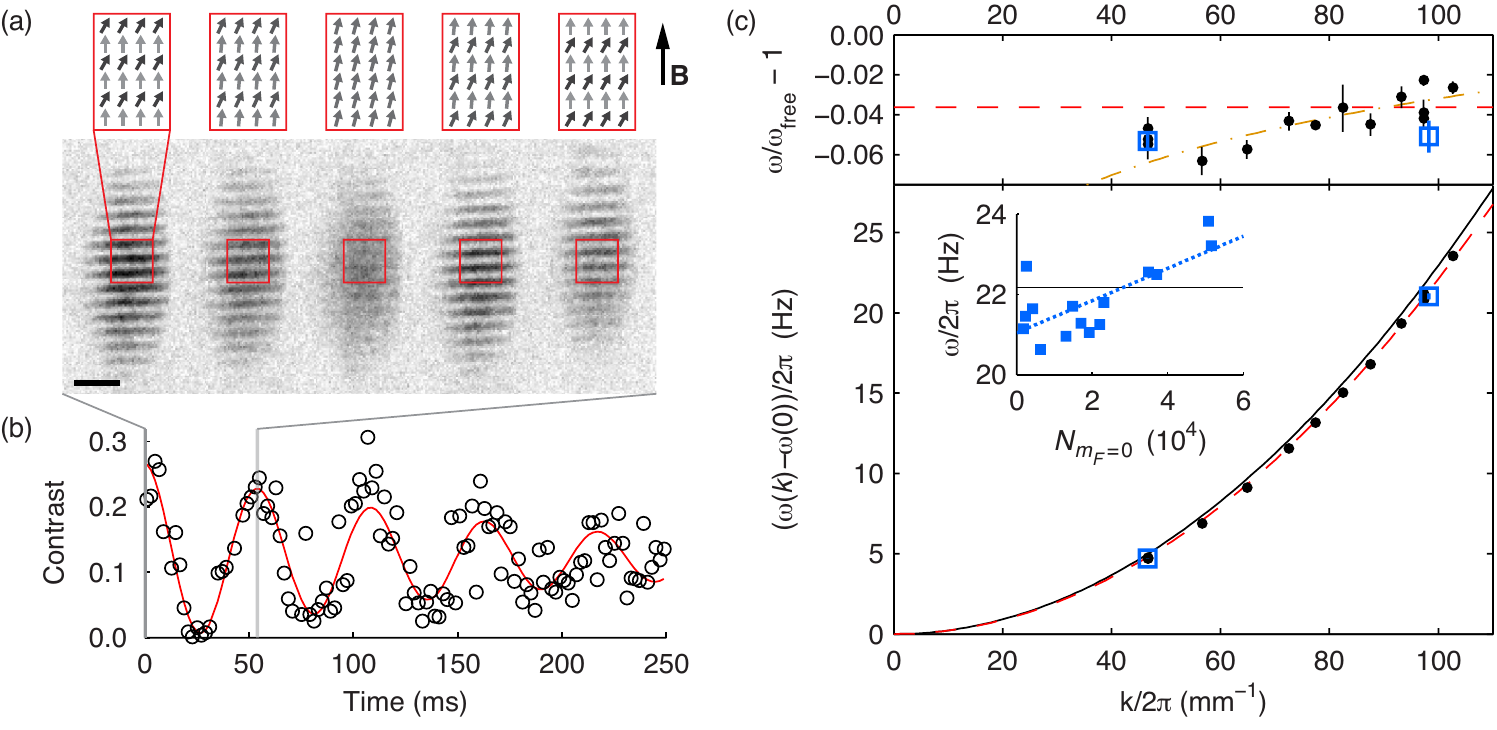}
\caption{Magnon dispersion relation measured with magnon contrast interferometry. (a) We initialize the system with a standing wave of magnons, in which the magnetization vector is periodically tilted in space (top row). For small tilt angles $\theta$, the magnon density is proportional to the atomic density in the $|m_F = 0\rangle$ hyperfine state, shown in images.  Here, $k = 2 \pi / (15.4\;\mu\mathrm{m})$ and the scalebar is $50\;\mu\mathrm{m}$. (b) From this initial state, the spatial modulation in the magnon density (open black circles) oscillates in time at frequency $2 (\omega(k)-\omega(0))$ (fit, solid red line). (c, bottom) The dispersion relation versus $k$.  Least-squares fit of the contrast oscillation frequency from single (filled black circles) or several values of $\theta$ (open blue squares) are corrected for frequency shifts due to the finite magnon populations (such frequency shift is shown in the inset, along with a linear fit).  The fitted quadratic dispersion relation (red dashed line) lies below the free-atom dispersion relation (solid black line).  (c, top) Fractional deviation of data from the free-atom result.  Data are fit to quadratic ($k^2$) (red dashed line) and power law ($k^\alpha$) models for $\alpha{=}2.04(1)_\mathrm{stat}$ (orange dot-dashed line). The error bars show $1 \, \sigma$ statistical uncertainty. \label{fig:dispersion}}
\end{center}
\end{figure*}

Our experiments begin with a $^{87}$Rb spinor Bose-Einstein condensate of about $10^6$ atoms held in an optical trap with trap frequencies $\omega_{x,y,z} = 2 \pi \times(4, 9, 300) \, \mbox{Hz}$, the strongest confinement being along the vertical $\mathbf{z}$ axis.  The atoms are initially polarized in the $|F{=}1, m_F{=}{-}1\rangle$ hyperfine state, where the quantization axis, lying in the plane of the prolate trap, is taken as that of the near-uniform 115 mG magnetic field applied to the gas.  Ferromagnetic ordering minimizes the free energy because the spin-dependent interaction $c_1^{(1)} n_0 = h \times 5.4\;\mathrm{Hz}$ is greater than twice the the quadratic Zeeman shift, $q = h\times 0.9\, \mbox{Hz}$ at the peak density of $n_0 = 1.5 \times 10^{14} \, \mbox{cm}^{-3}$\cite{Ho1998, Ohmi1998, Guzman2011}. The spin-dependent interaction $c_1^{(1)} = (4\pi\hbar^2/3m) (a_2-a_0)$ arises from the lower collisional energy between polarized spins (quantified by the scattering length $a_2$) than spins in the singlet state (quantified by the scattering length $a_0$).  We note that the presence of a non-zero, uniform magnetic field introduces a trivial gap to the magnon spectrum of $\Delta = \hbar \omega_L$, where $\omega_L = 2 \pi \times 80$ kHz is the Larmor frequency, but this gap is accounted for completely by considering the system in a rotating frame; in this frame, the magnons remain gapless\cite{Watanabe2013}.

We initialize the interferometer by imprinting a standing wave of magnons onto the condensate by performing spatially varying spin rotations of the gas. To do this, we briefly illuminate the atoms with two equal-frequency circularly polarized light beams directed nearly perpendicular to the plane of the trapped condensate and that intersect at the small relative angle $\vartheta$, creating a long wavelength intensity modulation along the condensate axis. The optical wavevector $k_L = 2 \pi (790.03 \, \mbox{nm})^{-1}$ is chosen so that the scalar ac Stark shift is zero, while the vector ac Stark shift acts akin to a transverse magnetic field \cite{Cohen-Tannoudji1972}.  The light intensity is modulated temporally at the Larmor frequency of $\omega_L = 2 \pi \times 80$ kHz, matching the resonance condition for a two-photon Raman transfer between Zeeman states. The varying light intensity causes the spins to be rotated in spin space by a polar angle $\theta(x,y) \propto I(x,y) \propto 1 + \cos(k y)$, where the local intensity of the light $I(x,y)$ has the form of a standing wave with wavevector $\mathbf{k} = \hat y\, 2 k_L \sin(\vartheta/2)$.  We assume here that the intensity of each beam, with $1/e^2$ radius of 300 $\mu$m, is constant and equal over the area of the condensate.

For weak excitation ($|\theta| \ll 1$), the resulting magnetization pattern is described as a ferromagnetic condensate excited with coherent populations of magnons at three wavevectors: $\mathbf{0}$, and $\pm \mathbf{k}$.  Allowing these populations to evolve with the magnon energies $E(0)$ and $E(k)$, respectively, the contrast (Fourier power) of the magnon density modulation at wavevector $k$ after a time $\tau$ is proportional to $\cos^2\left[(\omega(k)-\omega(0)) \tau\right]$ where $\hbar \omega(k) = E(k)$.  This temporal modulation thus measures the magnon dispersion relation minus the magnon energy gap.  As in Ref.\ \onlinecite{Gupta2002}, by obtaining our signal from the contrast of the interference fringes, and not from their phase, we remove errors due to the residual center-of-mass motion of the condensed gas in which the magnons evolve.

We detect the magnon density distribution using a new form of \emph{in situ} spin-dependent imaging (Fig.~\ref{fig:dispersion}).  In each of several consecutive images, we first apply a microwave pulse that resonantly transfers a fraction of atoms from one of the three Zeeman populations to the $F=2$ hyperfine level, and then perform absorption imaging with a short and intense pulse of light that propagates along $\mathbf{z}$ and is resonant with the $F=2 \rightarrow F^\prime = 3$ cycling transition on the D2 line.  The imaged atoms are rapidly expelled from the trap through light-induced forces, with no measurable impact on the remaining atoms in the $F=1$ hyperfine level.  These images give the in-trap column densities $\tilde{n}_{m_F}(x,y)$ for each of the Zeeman sublevels, integrated along the imaging axis. For $|\theta| \ll 1$, we typically employ a microwave $\pi$ pulse to image the entire magnon population, where the magnon column density is identical to $\tilde{n}_{0}(x,y)$.

Fig.~\ref{fig:dispersion} shows the spatially modulated magnon density as imprinted onto the condensate, and the subsequent temporal variation in the amplitude of that spatial modulation.  From these images, we quantify the evolution by calculating the power of the spatial Fourier component at wavevector $\mathbf{k}$.  We determine $\mathbf{k}$ by diverting the magnon-imprinting light onto a CCD camera with calibrated pixel sizes.  The angle of incidence onto the CCD matches that onto the plane of the condensate to within $10\;\mathrm{mrad}$.

To gather data more rapidly, we make twenty interferometric measurements on each ferromagnetic condensate. For each measurement, we create about $2\times 10^4$ magnons, image them after a time $\tau$, expel all remaining atoms in the $|F{=}1, m_F{=}0, {+}1\rangle$ states from the trap using a simultaneous microwave ($F{=}1 \to F{=}2$) and optical ($F{=}2\to F'{=}3$) pulse for $200\;\mu\mathrm{s}$, and then wait several hundred milliseconds to allow the remaining atoms to re-equilibrate.  This method of data collection allows us to identify and correct for any accidental quadrupolar collective excitation of the ferromagnetic condensate, which would otherwise shift the contrast modulation frequency.

At several wavevectors, we examined the variation of the observed contrast-modulation frequency with the strength of the magnon-imprinting optical pulse, i.e.\ as a function of the $m_F=0$ atom number ($N_{m_F=0}$).  We observe a frequency shift that varies roughly linearly with atom number, increasing by up to $\sim$2 Hz for $N_{m_F = 0} = 6\times10^4$ (1\% of the condensate number).  This frequency shift may be due to nonlinear effects on magnon propagation.  We adjust for this effect by fitting the observed variation and extrapolating all data to zero $m_F=0$ population.

Fitting the measured dispersion relation $\omega(k)$ (Fig.\ \ref{fig:dispersion}c) to the power-law form $\omega(k) = \hbar k^\alpha / 2 m^*$ gives $\alpha = 2.04(1)_\mathrm{stat}$, deviating from the expected quadratic dependence on momentum. This deviation could occur because we are probing at sufficiently large wavevectors that the quadratic approximation fails or because of an unresolved systematic error. We find the effective magnon mass (setting $\alpha=2$) to be $m^*{=}1.038(2)_\mathrm{stat}(8)_\mathrm{sys} \times m$.  This mass is significantly larger than the predictions of mean-field ($m^*/m =1$) and of beyond-mean-field calculations ($m^*/m = 1.003$) \cite{Phuc2013} based on $s$-wave spin-dependent interactions at zero temperature and at the peak density of our condensates.

\begin{figure}[tb!]
\begin{center}
\includegraphics[scale=1]{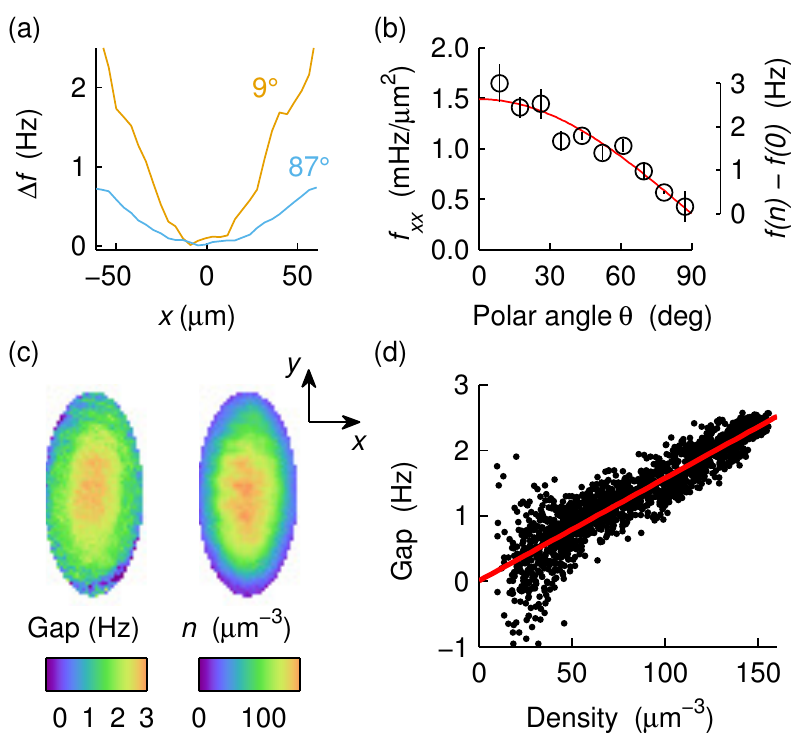}
\caption{The density-dependent energy shift $\Delta(n)$ of uniform ($k{=}0$) magnon excitations is measured by observing the Larmor precession of a ferromagnetic spinor condensate whose magnetization is tipped uniformly by a variable angle $\theta$ from the direction of an applied (in plane) magnetic field.  The spatially dependent Larmor precession frequency is deduced from measurements of one component of the transverse magnetization after $\tau = 100$ to $300$ ms of precession. (a) The frequency shift varies quadratically in position, with a curvature $f_{xx}=d^2f/dx^2$ (b) that varies with the cosine of the polar angle ($\theta$). We identify the curvature at $\theta = 90^\circ$ with the spatially inhomogeneous magnetic field. (c) A pixel-by-pixel analysis of the $\theta$-dependent frequency shift (left) shows a magnon energy gap map with the same geometry as the in-plane density of the condensate (right). The field of view is $125$ by $265\;\mu\mathrm{m}$. (d) A comparison of the gap and in-plane density of each pixel shows the expected linear trend. \label{fig:gap}}
\end{center}
\end{figure}

Magnetic dipolar interactions play an important role in our system because they break the rotational symmetry of spin interactions and open up a density-dependent gap $\Delta(n) = E(k=0)$. In an isotropic trap, Goldstone's Theorem predicts that the ferromagnetic state has gapless excitations because the spin interactions are symmetric. However, in our highly anisotropic trap, magnetic dipolar interactions create a spin-dependent energy that depends on the orientation of the spins with respect to the trap. 

For an infinite two-dimensional weakly dipolar gas with uniform magnetization that is tipped by the polar angle $\theta$ with respect to an in-plane magnetic field, the magnetic dipolar field changes the Larmor precession frequency by $\omega_\Delta = \frac{1}{\hbar} \Delta(n) \cos\theta$, with $\Delta(n) = \frac{2}{5} \mu_0 \mu^2 n$, where $\mu_0$ is the vacuum permeability, $\mu=g_F \mu_B$ is the atomic magnetic moment, $g_F=1/2$, $\mu_B$ is the Bohr magneton, $n$ is the peak density, and the prefactor accounts for a parabolic density distribution in the confined direction \cite{Kawaguchi2007}.  Noting that dipolar interactions are effectively short-ranged in two dimensions, we apply the local-density approximation and obtain a local value of the density-dependent magnon gap, using the above expression while replacing $n$ with the peak density at each imaged position $(x,y)$ in the gas.

To isolate this density-dependent shift from the effects of magnetic-field inhomogeneity, we characterize the spatial variation of the Larmor precession frequency in a manner similar to that demonstrated in Ref.~\onlinecite{Vengalattore2007}.  After preparing a longitudinally magnetized condensate, we apply a spatially uniform radiofrequency pulse that rotates the magnetization by the polar angle $\theta$.  Following $\tau = 100$--$300\;\mathrm{ms}$ of Larmor precession, we use our spin-dependent imaging first to determine the Zeeman populations along the magnetic field direction (to quantify $\theta$), and then, following a $\pi/2$ rf pulse, to characterize one transverse component of the precessed spins.  From the transverse spin image, we determine the spatially dependent Larmor precession phase $\Phi(x,y) = (\omega_B(x,y) + \omega_\Delta(x,y)) \tau$ which we treat as the sum of two terms: $\omega_B(x,y)$ determined by the spatially inhomogeneous magnetic field, and $\omega_\Delta(x,y)$.  The magnetic field is adequately described by its average value and spatial gradients, which vary between experimental runs, and spatial curvature, which is found to be constant between experimental repetitions.  The magnon energy gap is then determined at each imaged pixel by examining the variation of $\omega_\Delta(x,y)$ with tipping angle $\theta$.

We observe a density-dependent shift in the Larmor precession frequency (Fig.\ \ref{fig:gap}), indicating a magnon gap of $\Delta(n) = h \times 2.5(1)_\mathrm{stat}(2)_\mathrm{sys}$ Hz at the center of the condensate.  The systematic uncertainty is dominated by an uncertainty in the atomic density in our slightly anharmonic trapping potential. The measured value is consistent with our estimate of $\frac{2}{5} \mu_0 \mu^2 n_0 = h \times 2.35\;\mathrm{Hz}$, where we assume a harmonically trapped sample and calculate $n_0$ from the measured trap frequencies and total atom number.

This observed gap suggests an explanation for the unusually large value of $m^\ast$. The dispersion relation measurements are taken at low momentum, where the wavelength of 9--22$\;\mu\mathrm{m}$ is larger than the vertical Thomas-Fermi radius of the condensate $R_{TF} = 1.7\;\mu\mathrm{m}$ and the dipolar healing length $\hbar/\sqrt{\mu_0\mu^2 m n}\approx 4\;\mu\mathrm{m}$. At higher momentum, we would expect the dispersion relation to connect to that of a uniform three-dimensional spinor gas. Understanding the crossover in the magnon spectrum from (effectively) two- to three-dimensions merits theoretical work.


\begin{figure}[tb!]
\begin{center}
\includegraphics[scale=1]{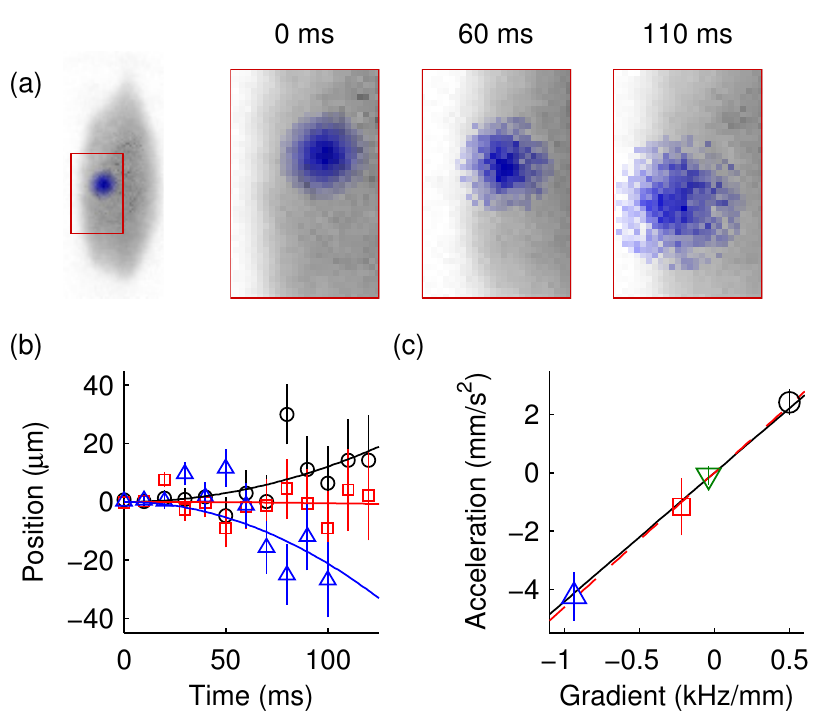}
\caption{Magnon magnetic moment. (a) Magnons are created as a small wavepacket (blue) within the larger condensate (gray). A magnetic field gradient accelerates the magnons, here to the lower-left.  The field gradient is independently measured with a Ramsey sequence to a third hyperfine state. The region of interest box is $52$ by $80\;\mu\mathrm{m}$. (b) Fits of position with time for three values of the magnetic field gradient. The measurement noise is heteroscedastic in time due to random center-of-mass motion of the condensate.  (c) The measured acceleration versus gradient (dashed red line) yields a magnetic moment consistent with the free-atom result (solid black line).  \label{fig:magnetic_moment}}
\end{center}
\end{figure}

Finally, we discuss our measurement of the magnon magnetic moment.  Neglecting interactions, a single magnon excitation, generated by infinitesimal rotation of a fully magnetized gas in the $|F{=}1, m_F{=}{-}1\rangle$ hyperfine state, is simply a single atom in the state $|F{=}1, m_F{=}0\rangle$.  This atom's magnetic moment has a zero-valued projection along the axis defined by the ferromagnetic order parameter.  However, measured against the ``magnon vacuum'' (the ferromagnet), the magnon magnetic moment $\mu^*$ is non-zero and, in the non-interacting limit, equal in magnitude to the atomic magnetic moment $\mu=g_F \mu_B$.

To measure $\mu^*$, we prepare a fully magnetized condensate in the presence of a magnetic field gradient of strength $B^\prime$, which displaces the equilibrium position of the condensate within the optical dipole trap. We then create a spatially localized magnon pulse, comprised of about $5000$ magnons, again using amplitude-modulated light with wavevector $k_L$, but modifying the optical setup so that a single light beam is focused to a small ($12 \, \mu m$ radius) spot.  Magnons created in the displaced condensate experience a chemical potential gradient, which leads to a differential force $\mu^* B^\prime$ between magnons and the ferromagnet in which they are excited. As expected, we observe that the magnon accelerates uniformly while the condensate remains still. From measurements of the magnon position versus elapsed time (Fig.\ \ref{fig:magnetic_moment}) we determine the magnon acceleration, and, therefrom, the ratio $\mu^* / m^*$.  We independently determine $B^\prime$ with a magnetometer sequence similar to the one used to determine the gap \footnote{A Ramsey microwave pulse sequence interferes atoms in a superposition of $|F{=}1, m_F{=}{-}1{\rangle}$ and $|F{=}2, m_F{=}{-}1{\rangle}$ hyperfine states. After a time $\tau$, the $F=2$ atomic density is spatially modulated at the wavevector $k = \Delta \mu \, B^\prime \tau/\hbar$ where $\Delta \mu$ is the difference in the magnetic moments of the two atomic states, which we assume to have the bare atomic values.}.

The ratio $\mu^*/m^* = -1.04(2)_\mathrm{stat}(8)\, g_F \mu_B/m$ is consistent with the mean-field prediction, where the mass $m^\ast$ is determined from magnon contrast interferometry. The 8\% systematic error, dominated by the uncertainty in the imaging magnification, lacks the resolution to confirm the heavy magnon mass observed above.

We have demonstrated a toolkit to create coherent magnons and measure their basic properties in an ultracold Bose gas. While the measured magnetic moment and many-body gap are close to the mean-field theoretical expectations, we find that the magnon mass is substantially heavier. The accuracy of the measurement provides a challenge to theoretical calculations. Moreover, the techniques demonstrated here can be applied to other intriguing ultracold systems, including the non-ferromagnetic spinor gases and dipolar gases.

We thank M. Solarz for help designing and building the cold atom setup and J. Guzman and G. B. Jo for help implementing a spinor gas experiment. This work was supported by DTRA (Contract No.\ HDTRA1-09-1-0020) and by the Army Research Office with funding from the DARPA Optical Lattice Emulator program.  G.E.M.\ acknowledges support from the Hertz Foundation.
	
\bibliography{magnons}

\begin{thebibliography}{22}%
\makeatletter
\providecommand \@ifxundefined [1]{%
 \@ifx{#1\undefined}
}%
\providecommand \@ifnum [1]{%
 \ifnum #1\expandafter \@firstoftwo
 \else \expandafter \@secondoftwo
 \fi
}%
\providecommand \@ifx [1]{%
 \ifx #1\expandafter \@firstoftwo
 \else \expandafter \@secondoftwo
 \fi
}%
\providecommand \natexlab [1]{#1}%
\providecommand \enquote  [1]{``#1''}%
\providecommand \bibnamefont  [1]{#1}%
\providecommand \bibfnamefont [1]{#1}%
\providecommand \citenamefont [1]{#1}%
\providecommand \href@noop [0]{\@secondoftwo}%
\providecommand \href [0]{\begingroup \@sanitize@url \@href}%
\providecommand \@href[1]{\@@startlink{#1}\@@href}%
\providecommand \@@href[1]{\endgroup#1\@@endlink}%
\providecommand \@sanitize@url [0]{\catcode `\\12\catcode `\$12\catcode
  `\&12\catcode `\#12\catcode `\^12\catcode `\_12\catcode `\%12\relax}%
\providecommand \@@startlink[1]{}%
\providecommand \@@endlink[0]{}%
\providecommand \url  [0]{\begingroup\@sanitize@url \@url }%
\providecommand \@url [1]{\endgroup\@href {#1}{\urlprefix }}%
\providecommand \urlprefix  [0]{URL }%
\providecommand \Eprint [0]{\href }%
\providecommand \doibase [0]{http://dx.doi.org/}%
\providecommand \selectlanguage [0]{\@gobble}%
\providecommand \bibinfo  [0]{\@secondoftwo}%
\providecommand \bibfield  [0]{\@secondoftwo}%
\providecommand \translation [1]{[#1]}%
\providecommand \BibitemOpen [0]{}%
\providecommand \bibitemStop [0]{}%
\providecommand \bibitemNoStop [0]{.\EOS\space}%
\providecommand \EOS [0]{\spacefactor3000\relax}%
\providecommand \BibitemShut  [1]{\csname bibitem#1\endcsname}%
\let\auto@bib@innerbib\@empty
\bibitem [{\citenamefont {{G}uzman}\ \emph {et~al.}(2011)\citenamefont
  {{G}uzman}, \citenamefont {{J}o}, \citenamefont {{W}enz}, \citenamefont
  {{M}urch}, \citenamefont {{T}homas},\ and\ \citenamefont
  {{S}tamper{-}{K}urn}}]{Guzman2011}%
  \BibitemOpen
  \bibfield  {author} {\bibinfo {author} {\bibfnamefont {J.}~\bibnamefont
  {{G}uzman}}, \bibinfo {author} {\bibfnamefont {G.-B.}\ \bibnamefont {{J}o}},
  \bibinfo {author} {\bibfnamefont {A.~N.}\ \bibnamefont {{W}enz}}, \bibinfo
  {author} {\bibfnamefont {K.~W.}\ \bibnamefont {{M}urch}}, \bibinfo {author}
  {\bibfnamefont {C.~K.}\ \bibnamefont {{T}homas}}, \ and\ \bibinfo {author}
  {\bibfnamefont {D.~M.}\ \bibnamefont {{S}tamper{-}{K}urn}},\ }\href {\doibase
  10.1103/PhysRevA.84.063625} {\bibfield  {journal} {\bibinfo  {journal}
  {{P}hys. {R}ev. {A}}\ }\textbf {\bibinfo {volume} {84}},\ \bibinfo {pages}
  {063625} (\bibinfo {year} {2011})}\BibitemShut {NoStop}%
\bibitem [{\citenamefont {{G}reif}\ \emph {et~al.}(2013)\citenamefont
  {{G}reif}, \citenamefont {{U}ehlinger}, \citenamefont {{J}otzu},
  \citenamefont {{T}arruell},\ and\ \citenamefont {{E}sslinger}}]{Greif2013}%
  \BibitemOpen
  \bibfield  {author} {\bibinfo {author} {\bibfnamefont {D.}~\bibnamefont
  {{G}reif}}, \bibinfo {author} {\bibfnamefont {T.}~\bibnamefont
  {{U}ehlinger}}, \bibinfo {author} {\bibfnamefont {G.}~\bibnamefont
  {{J}otzu}}, \bibinfo {author} {\bibfnamefont {L.}~\bibnamefont {{T}arruell}},
  \ and\ \bibinfo {author} {\bibfnamefont {T.}~\bibnamefont {{E}sslinger}},\
  }\href {\doibase 10.1126/science.1236362} {\bibfield  {journal} {\bibinfo
  {journal} {{S}cience}\ }\textbf {\bibinfo {volume} {340}},\ \bibinfo {pages}
  {1307} (\bibinfo {year} {2013})},\ \Eprint
  {http://arxiv.org/abs/http://www.sciencemag.org/content/340/6138/1307.full.pdf}
  {http://www.sciencemag.org/content/340/6138/1307.full.pdf} \BibitemShut
  {NoStop}%
\bibitem [{\citenamefont {{I}slam}\ \emph {et~al.}(2013)\citenamefont
  {{I}slam}, \citenamefont {{S}enko}, \citenamefont {{C}ampbell}, \citenamefont
  {{K}orenblit}, \citenamefont {{S}mith}, \citenamefont {{L}ee}, \citenamefont
  {{E}dwards}, \citenamefont {{W}ang}, \citenamefont {{F}reericks},\ and\
  \citenamefont {{M}onroe}}]{Islam2013}%
  \BibitemOpen
  \bibfield  {author} {\bibinfo {author} {\bibfnamefont {R.}~\bibnamefont
  {{I}slam}}, \bibinfo {author} {\bibfnamefont {C.}~\bibnamefont {{S}enko}},
  \bibinfo {author} {\bibfnamefont {W.~C.}\ \bibnamefont {{C}ampbell}},
  \bibinfo {author} {\bibfnamefont {S.}~\bibnamefont {{K}orenblit}}, \bibinfo
  {author} {\bibfnamefont {J.}~\bibnamefont {{S}mith}}, \bibinfo {author}
  {\bibfnamefont {A.}~\bibnamefont {{L}ee}}, \bibinfo {author} {\bibfnamefont
  {E.~E.}\ \bibnamefont {{E}dwards}}, \bibinfo {author} {\bibfnamefont
  {C.-C.~J.}\ \bibnamefont {{W}ang}}, \bibinfo {author} {\bibfnamefont {J.~K.}\
  \bibnamefont {{F}reericks}}, \ and\ \bibinfo {author} {\bibfnamefont
  {C.}~\bibnamefont {{M}onroe}},\ }\href {\doibase 10.1126/science.1232296}
  {\bibfield  {journal} {\bibinfo  {journal} {{S}cience}\ }\textbf {\bibinfo
  {volume} {340}},\ \bibinfo {pages} {583} (\bibinfo {year} {2013})},\ \Eprint
  {http://arxiv.org/abs/http://www.sciencemag.org/content/340/6132/583.full.pdf}
  {http://www.sciencemag.org/content/340/6132/583.full.pdf} \BibitemShut
  {NoStop}%
\bibitem [{\citenamefont {{S}tamper {K}urn}\ and\ \citenamefont
  {{U}eda}(2013)}]{Stamper-Kurn2013}%
  \BibitemOpen
  \bibfield  {author} {\bibinfo {author} {\bibfnamefont {D.~M.}\ \bibnamefont
  {{S}tamper {K}urn}}\ and\ \bibinfo {author} {\bibfnamefont {M.}~\bibnamefont
  {{U}eda}},\ }\href {\doibase 10.1103/RevModPhys.85.1191} {\bibfield
  {journal} {\bibinfo  {journal} {{R}ev. {M}od. {P}hys.}\ }\textbf {\bibinfo
  {volume} {85}},\ \bibinfo {pages} {1191} (\bibinfo {year}
  {2013})}\BibitemShut {NoStop}%
\bibitem [{\citenamefont {{G}orshkov}\ \emph {et~al.}(2010)\citenamefont
  {{G}orshkov}, \citenamefont {{H}ermele}, \citenamefont {{G}urarie},
  \citenamefont {{X}u}, \citenamefont {{J}ulienne}, \citenamefont {{Y}e},
  \citenamefont {{Z}oller}, \citenamefont {{D}emler}, \citenamefont {{L}ukin},\
  and\ \citenamefont {{R}ey}}]{Gorshkov2010}%
  \BibitemOpen
  \bibfield  {author} {\bibinfo {author} {\bibfnamefont {A.~V.}\ \bibnamefont
  {{G}orshkov}}, \bibinfo {author} {\bibfnamefont {M.}~\bibnamefont
  {{H}ermele}}, \bibinfo {author} {\bibfnamefont {V.}~\bibnamefont
  {{G}urarie}}, \bibinfo {author} {\bibfnamefont {C.}~\bibnamefont {{X}u}},
  \bibinfo {author} {\bibfnamefont {P.~S.}\ \bibnamefont {{J}ulienne}},
  \bibinfo {author} {\bibfnamefont {J.}~\bibnamefont {{Y}e}}, \bibinfo {author}
  {\bibfnamefont {P.}~\bibnamefont {{Z}oller}}, \bibinfo {author}
  {\bibfnamefont {E.}~\bibnamefont {{D}emler}}, \bibinfo {author}
  {\bibfnamefont {M.~D.}\ \bibnamefont {{L}ukin}}, \ and\ \bibinfo {author}
  {\bibfnamefont {A.~M.}\ \bibnamefont {{R}ey}},\ }\href
  {http://dx.doi.org/10.1038/nphys1535} {\bibfield  {journal} {\bibinfo
  {journal} {{N}at {P}hys}\ }\textbf {\bibinfo {volume} {6}},\ \bibinfo {pages}
  {289} (\bibinfo {year} {2010})}\BibitemShut {NoStop}%
\bibitem [{\citenamefont {{Z}hang}\ \emph {et~al.}(2014)\citenamefont
  {{Z}hang}, \citenamefont {{B}ishof}, \citenamefont {{B}romley}, \citenamefont
  {{K}raus}, \citenamefont {{S}afronova}, \citenamefont {{Z}oller},
  \citenamefont {{R}ey},\ and\ \citenamefont {{Y}e}}]{Zhang2014}%
  \BibitemOpen
  \bibfield  {author} {\bibinfo {author} {\bibfnamefont {X.}~\bibnamefont
  {{Z}hang}}, \bibinfo {author} {\bibfnamefont {M.}~\bibnamefont {{B}ishof}},
  \bibinfo {author} {\bibfnamefont {S.~L.}\ \bibnamefont {{B}romley}}, \bibinfo
  {author} {\bibfnamefont {C.~V.}\ \bibnamefont {{K}raus}}, \bibinfo {author}
  {\bibfnamefont {M.~S.}\ \bibnamefont {{S}afronova}}, \bibinfo {author}
  {\bibfnamefont {P.}~\bibnamefont {{Z}oller}}, \bibinfo {author}
  {\bibfnamefont {A.~M.}\ \bibnamefont {{R}ey}}, \ and\ \bibinfo {author}
  {\bibfnamefont {J.}~\bibnamefont {{Y}e}},\ }\href@noop {} {\bibfield
  {journal} {\bibinfo  {journal} {ar{X}iv}\ } (\bibinfo {year} {2014})},\
  \Eprint {http://arxiv.org/abs/1403.2964} {1403.2964} \BibitemShut {NoStop}%
\bibitem [{\citenamefont {{S}cazza}\ \emph {et~al.}(2014)\citenamefont
  {{S}cazza}, \citenamefont {{H}ofrichter}, \citenamefont {{H}öfer},
  \citenamefont {{D}e {G}root}, \citenamefont {{B}loch},\ and\ \citenamefont
  {{F}ölling}}]{Scazza2014}%
  \BibitemOpen
  \bibfield  {author} {\bibinfo {author} {\bibfnamefont {F.}~\bibnamefont
  {{S}cazza}}, \bibinfo {author} {\bibfnamefont {C.}~\bibnamefont
  {{H}ofrichter}}, \bibinfo {author} {\bibfnamefont {M.}~\bibnamefont
  {{H}öfer}}, \bibinfo {author} {\bibfnamefont {P.~C.}\ \bibnamefont {{D}e
  {G}root}}, \bibinfo {author} {\bibfnamefont {I.}~\bibnamefont {{B}loch}}, \
  and\ \bibinfo {author} {\bibfnamefont {S.}~\bibnamefont {{F}ölling}},\
  }\href@noop {} {\bibfield  {journal} {\bibinfo  {journal} {ar{X}iv}\ }
  (\bibinfo {year} {2014})},\ \Eprint {http://arxiv.org/abs/1403.4761}
  {1403.4761} \BibitemShut {NoStop}%
\bibitem [{\citenamefont {{M}icheli}\ \emph {et~al.}(2006)\citenamefont
  {{M}icheli}, \citenamefont {{B}rennen},\ and\ \citenamefont
  {{Z}oller}}]{Micheli2006}%
  \BibitemOpen
  \bibfield  {author} {\bibinfo {author} {\bibfnamefont {A.}~\bibnamefont
  {{M}icheli}}, \bibinfo {author} {\bibfnamefont {G.~K.}\ \bibnamefont
  {{B}rennen}}, \ and\ \bibinfo {author} {\bibfnamefont {P.}~\bibnamefont
  {{Z}oller}},\ }\href {http://dx.doi.org/10.1038/nphys287} {\bibfield
  {journal} {\bibinfo  {journal} {{N}at {P}hys}\ }\textbf {\bibinfo {volume}
  {2}},\ \bibinfo {pages} {341} (\bibinfo {year} {2006})}\BibitemShut {NoStop}%
\bibitem [{\citenamefont {{S}adler}\ \emph {et~al.}(2006)\citenamefont
  {{S}adler}, \citenamefont {{H}igbie}, \citenamefont {{L}eslie}, \citenamefont
  {{V}engalattore},\ and\ \citenamefont {{S}tamper {K}urn}}]{Sadler2006}%
  \BibitemOpen
  \bibfield  {author} {\bibinfo {author} {\bibfnamefont {L.~E.}\ \bibnamefont
  {{S}adler}}, \bibinfo {author} {\bibfnamefont {J.~M.}\ \bibnamefont
  {{H}igbie}}, \bibinfo {author} {\bibfnamefont {S.~R.}\ \bibnamefont
  {{L}eslie}}, \bibinfo {author} {\bibfnamefont {M.}~\bibnamefont
  {{V}engalattore}}, \ and\ \bibinfo {author} {\bibfnamefont {D.~M.}\
  \bibnamefont {{S}tamper {K}urn}},\ }\href
  {http://dx.doi.org/10.1038/nature05094} {\bibfield  {journal} {\bibinfo
  {journal} {{N}ature}\ }\textbf {\bibinfo {volume} {443}},\ \bibinfo {pages}
  {312} (\bibinfo {year} {2006})}\BibitemShut {NoStop}%
\bibitem [{\citenamefont {{C}hang}\ \emph {et~al.}(2005)\citenamefont
  {{C}hang}, \citenamefont {{Q}in}, \citenamefont {{Z}hang}, \citenamefont
  {{Y}ou},\ and\ \citenamefont {{C}hapman}}]{Chang2005}%
  \BibitemOpen
  \bibfield  {author} {\bibinfo {author} {\bibfnamefont {M.-S.}\ \bibnamefont
  {{C}hang}}, \bibinfo {author} {\bibfnamefont {Q.}~\bibnamefont {{Q}in}},
  \bibinfo {author} {\bibfnamefont {W.}~\bibnamefont {{Z}hang}}, \bibinfo
  {author} {\bibfnamefont {L.}~\bibnamefont {{Y}ou}}, \ and\ \bibinfo {author}
  {\bibfnamefont {M.~S.}\ \bibnamefont {{C}hapman}},\ }\href
  {http://dx.doi.org/10.1038/nphys153} {\bibfield  {journal} {\bibinfo
  {journal} {{N}at {P}hys}\ }\textbf {\bibinfo {volume} {1}},\ \bibinfo {pages}
  {111} (\bibinfo {year} {2005})}\BibitemShut {NoStop}%
\bibitem [{\citenamefont {{L}iu}\ \emph {et~al.}(2009)\citenamefont {{L}iu},
  \citenamefont {{J}ung}, \citenamefont {{M}axwell}, \citenamefont {{T}urner},
  \citenamefont {{T}iesinga},\ and\ \citenamefont {{L}ett}}]{Liu2009}%
  \BibitemOpen
  \bibfield  {author} {\bibinfo {author} {\bibfnamefont {Y.}~\bibnamefont
  {{L}iu}}, \bibinfo {author} {\bibfnamefont {S.}~\bibnamefont {{J}ung}},
  \bibinfo {author} {\bibfnamefont {S.~E.}\ \bibnamefont {{M}axwell}}, \bibinfo
  {author} {\bibfnamefont {L.~D.}\ \bibnamefont {{T}urner}}, \bibinfo {author}
  {\bibfnamefont {E.}~\bibnamefont {{T}iesinga}}, \ and\ \bibinfo {author}
  {\bibfnamefont {P.~D.}\ \bibnamefont {{L}ett}},\ }\href {\doibase
  10.1103/PhysRevLett.102.125301} {\bibfield  {journal} {\bibinfo  {journal}
  {{P}hys. {R}ev. {L}ett.}\ }\textbf {\bibinfo {volume} {102}},\ \bibinfo
  {pages} {125301} (\bibinfo {year} {2009})}\BibitemShut {NoStop}%
\bibitem [{\citenamefont {{F}ukuhara}\ \emph {et~al.}(2013)\citenamefont
  {{F}ukuhara}, \citenamefont {{S}chausz}, \citenamefont {{E}ndres},
  \citenamefont {{H}ild}, \citenamefont {{C}heneau}, \citenamefont {{B}loch},\
  and\ \citenamefont {{G}ross}}]{Fukuhara2013}%
  \BibitemOpen
  \bibfield  {author} {\bibinfo {author} {\bibfnamefont {T.}~\bibnamefont
  {{F}ukuhara}}, \bibinfo {author} {\bibfnamefont {P.}~\bibnamefont
  {{S}chausz}}, \bibinfo {author} {\bibfnamefont {M.}~\bibnamefont {{E}ndres}},
  \bibinfo {author} {\bibfnamefont {S.}~\bibnamefont {{H}ild}}, \bibinfo
  {author} {\bibfnamefont {M.}~\bibnamefont {{C}heneau}}, \bibinfo {author}
  {\bibfnamefont {I.}~\bibnamefont {{B}loch}}, \ and\ \bibinfo {author}
  {\bibfnamefont {C.}~\bibnamefont {{G}ross}},\ }\href
  {http://dx.doi.org/10.1038/nature12541} {\bibfield  {journal} {\bibinfo
  {journal} {{N}ature}\ }\textbf {\bibinfo {volume} {502}},\ \bibinfo {pages}
  {76} (\bibinfo {year} {2013})}\BibitemShut {NoStop}%
\bibitem [{\citenamefont {{H}o}(1998)}]{Ho1998}%
  \BibitemOpen
  \bibfield  {author} {\bibinfo {author} {\bibfnamefont {T.-L.}\ \bibnamefont
  {{H}o}},\ }\href {\doibase 10.1103/PhysRevLett.81.742} {\bibfield  {journal}
  {\bibinfo  {journal} {{P}hys. {R}ev. {L}ett.}\ }\textbf {\bibinfo {volume}
  {81}},\ \bibinfo {pages} {742} (\bibinfo {year} {1998})}\BibitemShut
  {NoStop}%
\bibitem [{\citenamefont {{T}etsuo {O}hmi}\ and\ \citenamefont {{K}azushige
  {M}achida}(1998)}]{Ohmi1998}%
  \BibitemOpen
  \bibfield  {author} {\bibinfo {author} {\bibnamefont {{T}etsuo {O}hmi}}\ and\
  \bibinfo {author} {\bibnamefont {{K}azushige {M}achida}},\ }\href {\doibase
  10.1143/JPSJ.67.1822} {\bibfield  {journal} {\bibinfo  {journal} {{J}ournal
  of the {P}hysical {S}ociety of {J}apan}\ }\textbf {\bibinfo {volume} {67}},\
  \bibinfo {pages} {1822} (\bibinfo {year} {1998})}\BibitemShut {NoStop}%
\bibitem [{\citenamefont {{W}atanabe}\ and\ \citenamefont
  {{M}urayama}(2012)}]{Watanabe2012}%
  \BibitemOpen
  \bibfield  {author} {\bibinfo {author} {\bibfnamefont {H.}~\bibnamefont
  {{W}atanabe}}\ and\ \bibinfo {author} {\bibfnamefont {H.}~\bibnamefont
  {{M}urayama}},\ }\href {\doibase 10.1103/PhysRevLett.108.251602} {\bibfield
  {journal} {\bibinfo  {journal} {{P}hys. {R}ev. {L}ett.}\ }\textbf {\bibinfo
  {volume} {108}},\ \bibinfo {pages} {251602} (\bibinfo {year}
  {2012})}\BibitemShut {NoStop}%
\bibitem [{\citenamefont {{G}upta}\ \emph {et~al.}(2002)\citenamefont
  {{G}upta}, \citenamefont {{D}ieckmann}, \citenamefont {{H}adzibabic},\ and\
  \citenamefont {{P}ritchard}}]{Gupta2002}%
  \BibitemOpen
  \bibfield  {author} {\bibinfo {author} {\bibfnamefont {S.}~\bibnamefont
  {{G}upta}}, \bibinfo {author} {\bibfnamefont {K.}~\bibnamefont
  {{D}ieckmann}}, \bibinfo {author} {\bibfnamefont {Z.}~\bibnamefont
  {{H}adzibabic}}, \ and\ \bibinfo {author} {\bibfnamefont {D.~E.}\
  \bibnamefont {{P}ritchard}},\ }\href {\doibase 10.1103/PhysRevLett.89.140401}
  {\bibfield  {journal} {\bibinfo  {journal} {{P}hys. {R}ev. {L}ett.}\ }\textbf
  {\bibinfo {volume} {89}},\ \bibinfo {pages} {140401} (\bibinfo {year}
  {2002})}\BibitemShut {NoStop}%
\bibitem [{\citenamefont {{N}guyen~{T}hanh {P}huc}\ \emph
  {et~al.}(2013)\citenamefont {{N}guyen~{T}hanh {P}huc}, \citenamefont {{Y}uki
  {K}awaguchi},\ and\ \citenamefont {{M}asahito {U}eda}}]{Phuc2013}%
  \BibitemOpen
  \bibfield  {author} {\bibinfo {author} {\bibnamefont {{N}guyen~{T}hanh
  {P}huc}}, \bibinfo {author} {\bibnamefont {{Y}uki {K}awaguchi}}, \ and\
  \bibinfo {author} {\bibnamefont {{M}asahito {U}eda}},\ }\href {\doibase
  http://dx.doi.org/10.1016/j.aop.2012.10.004} {\bibfield  {journal} {\bibinfo
  {journal} {{A}nnals of {P}hysics}\ }\textbf {\bibinfo {volume} {328}},\
  \bibinfo {pages} {158 } (\bibinfo {year} {2013})}\BibitemShut {NoStop}%
\bibitem [{\citenamefont {{W}atanabe}\ \emph {et~al.}(2013)\citenamefont
  {{W}atanabe}, \citenamefont {{B}rauner},\ and\ \citenamefont
  {{M}urayama}}]{Watanabe2013}%
  \BibitemOpen
  \bibfield  {author} {\bibinfo {author} {\bibfnamefont {H.}~\bibnamefont
  {{W}atanabe}}, \bibinfo {author} {\bibfnamefont {T.}~\bibnamefont
  {{B}rauner}}, \ and\ \bibinfo {author} {\bibfnamefont {H.}~\bibnamefont
  {{M}urayama}},\ }\href {\doibase 10.1103/PhysRevLett.111.021601} {\bibfield
  {journal} {\bibinfo  {journal} {{P}hys. {R}ev. {L}ett.}\ }\textbf {\bibinfo
  {volume} {111}},\ \bibinfo {pages} {021601} (\bibinfo {year}
  {2013})}\BibitemShut {NoStop}%
\bibitem [{\citenamefont {{C}ohen {T}annoudji}\ and\ \citenamefont {{D}upont
  {R}oc}(1972)}]{Cohen-Tannoudji1972}%
  \BibitemOpen
  \bibfield  {author} {\bibinfo {author} {\bibfnamefont {C.}~\bibnamefont
  {{C}ohen {T}annoudji}}\ and\ \bibinfo {author} {\bibfnamefont
  {J.}~\bibnamefont {{D}upont {R}oc}},\ }\href {\doibase
  10.1103/PhysRevA.5.968} {\bibfield  {journal} {\bibinfo  {journal} {{P}hys.
  {R}ev. {A}}\ }\textbf {\bibinfo {volume} {5}},\ \bibinfo {pages} {968}
  (\bibinfo {year} {1972})}\BibitemShut {NoStop}%
\bibitem [{\citenamefont {{K}awaguchi}\ \emph {et~al.}(2007)\citenamefont
  {{K}awaguchi}, \citenamefont {{S}aito},\ and\ \citenamefont
  {{U}eda}}]{Kawaguchi2007}%
  \BibitemOpen
  \bibfield  {author} {\bibinfo {author} {\bibfnamefont {Y.}~\bibnamefont
  {{K}awaguchi}}, \bibinfo {author} {\bibfnamefont {H.}~\bibnamefont
  {{S}aito}}, \ and\ \bibinfo {author} {\bibfnamefont {M.}~\bibnamefont
  {{U}eda}},\ }\href {\doibase 10.1103/PhysRevLett.98.110406} {\bibfield
  {journal} {\bibinfo  {journal} {{P}hys. {R}ev. {L}ett.}\ }\textbf {\bibinfo
  {volume} {98}},\ \bibinfo {pages} {110406} (\bibinfo {year}
  {2007})}\BibitemShut {NoStop}%
\bibitem [{\citenamefont {{V}engalattore}\ \emph {et~al.}(2007)\citenamefont
  {{V}engalattore}, \citenamefont {{H}igbie}, \citenamefont {{L}eslie},
  \citenamefont {{G}uzman}, \citenamefont {{S}adler},\ and\ \citenamefont
  {{S}tamper{-}{K}urn}}]{Vengalattore2007}%
  \BibitemOpen
  \bibfield  {author} {\bibinfo {author} {\bibfnamefont {M.}~\bibnamefont
  {{V}engalattore}}, \bibinfo {author} {\bibfnamefont {J.~M.}\ \bibnamefont
  {{H}igbie}}, \bibinfo {author} {\bibfnamefont {S.~R.}\ \bibnamefont
  {{L}eslie}}, \bibinfo {author} {\bibfnamefont {J.}~\bibnamefont {{G}uzman}},
  \bibinfo {author} {\bibfnamefont {L.~E.}\ \bibnamefont {{S}adler}}, \ and\
  \bibinfo {author} {\bibfnamefont {D.~M.}\ \bibnamefont
  {{S}tamper{-}{K}urn}},\ }\href {http://link.aps.org/abstract/PRL/v98/e200801}
  {\bibfield  {journal} {\bibinfo  {journal} {{P}hysical {R}eview {L}etters}\
  }\textbf {\bibinfo {volume} {98}},\ \bibinfo {eid} {200801} (\bibinfo {year}
  {2007})}\BibitemShut {NoStop}%
\bibitem [{Note1()}]{Note1}%
  \BibitemOpen
  \bibinfo {note} {A Ramsey microwave pulse sequence interferes atoms in a
  superposition of $|F{=}1, m_F{=}{-}1{\delimiter "526930B }$ and $|F{=}2,
  m_F{=}{-}1{\delimiter "526930B }$ hyperfine states. After a time $\tau $, the
  $F=2$ atomic density is spatially modulated at the wavevector $k = \Delta \mu
  \protect \tmspace +\thinmuskip {.1667em} B^\prime \tau /\hbar $ where $\Delta
  \mu $ is the difference in the magnetic moments of the two atomic states,
  which we assume to have the bare atomic values.}\BibitemShut {Stop}%
\end{thebibliography}%

\clearpage

\end{document}